\documentclass[prb,twocolumn]{revtex4-1}

\usepackage{amssymb}
\usepackage{amsmath}
\usepackage[dvips]{graphicx}

\begin{document}

\title{Magnetostatic bias in multilayer microwires: theory and experiments}
\author{J. Escrig, S. Allende and D. Altbir}
\author{M. Bahiana}
\author{J. Torrej\'on, G. Badini and M. V\'azquez}
\affiliation{Departamento de F\'{\i}sica, Universidad de Santiago de Chile (USACH), Av.
Ecuador 3493, 917-0124 Santiago, Chile.}

\affiliation{Universidade Federal do Rio de Janeiro, Instituto de F\'{\i}sica, CP 68528,
21941-972 Rio de Janeiro, Brazil}

\affiliation{Instituto de Ciencia de Materiales-CSIC, Campus de Cantoblanco, 28049
Madrid, Spain.}


\begin{abstract}
The hysteresis curves of multilayer microwires consisting of a soft magnetic
nucleus, intermediate non-magnetic layers, and an external hard magnetic
layer are investigated. The magnetostatic interaction between magnetic
layers is proved to give rise to an antiferromagnetic-like coupling
resulting in a magnetostatic bias in the hysteresis curves of the soft
nucleus. This magnetostatic biasing effect is investigated in terms of the
microwire geometry. The experimental results are interpreted considering an
analytical model taking into account the magnetostatic interaction between
the magnetic layers.
\end{abstract}

\maketitle

\section{Introduction}

During the last decade, soft magnetic materials have been deeply
investigated. Besides the basic scientific interest in their magnetic
properties, there is a great deal of technologic interest due to their use
in sensing applications, particularly in the fields of automotive, mobile
communication, medical and home appliance industries. \cite{Vazquez07,
CTD+04, MH07, KMP03} Moreover, these materials are very promising for
spintronic devices in magnetic recording media. \cite{YKN+07} Two types of
soft magnetic microwire families are currently studied: in-water quenched
amorphous wires with diameters of around 120 $\mu $m, \cite{OU95} and
quenched and drawn microwires with diameters ranging from around 2 to 20 $%
\mu $m, \cite{ZGV+03} covered by a protective insulating glassy coat. The
most interesting magnetic property observed in these ultrasoft magnetic
microwires is a single Barkhausen jump that results in a square-shaped
hysteresis loop found in Fe-based microwires, as a consequence of
magnetization reversal by displacement of a single domain wall. \cite{VGV05}
The development of preparation techniques had lead to the production of well
controlled multilayered systems composed by layers of different materials.
In particular, magnetic multilayers have been grown, like
magnetic/nonmagnetic trilayers and multisegmented wires. Virtually all the
multilayer magnetic systems present magnetostatic bias effects as a
consequence of the coupling between adjacent layers with different magnetic
character. In such systems, the low-field hysteresis loops are shifted along
the field direction by an amount which is labeled as magnetostatic bias
field, $H_{b}$. In fact, the magnetostatic bias field acts as an additional
magnetic field on the layer that magnetizes at lower fields, being
equivalent to an unidirectional magnetic anisotropy. The role played by
magnetic biasing in controlling the magnetization reversal in
heterostructures is a key issue for applications related to spin electronic
devices and ultrahigh density recording media. In fact, this is in the base
of spin-valve and magnetic tunnel junction's heads where magnetic exchange
coupling between ferromagnetic (FM) and antiferromagnetic (AFM) layered
structures facilitates the material to exhibit a well defined non-symmetric
response to magnetic excitations. \cite{NS99}

In the last years, a family of multilayer microwires has been introduced by
Pirota \textit{et al}. \cite{PHN+04} in which quenching and drawing,
sputtering and electroplating techniques have been combined to prepare
multilayer microwires consisting of two metallic layers separated by an
intermediate insulating microlayer: a micrometric cylindrical nucleus with a
radius $R_{i}$ and an external metallic outer microtube with internal and
external radii $a_{e}$ and $R_{e}$, respectively. The length of the
microwire is $L$. In order to simplify the illustration of our results, we
present figures in terms of $R_{i}$, $t_{s}$ and $t_{e}$, where $%
t_{s}=a_{e}-R_{i}$ is the thickness of the intermediate non-magnetic spacer
layer and $t_{e}=R_{e}-a_{e}$ represents the thickness of the external
microtube, as illustrated in Fig. 1(a).

\begin{figure}[h]
\begin{center}
\includegraphics[width=8cm]{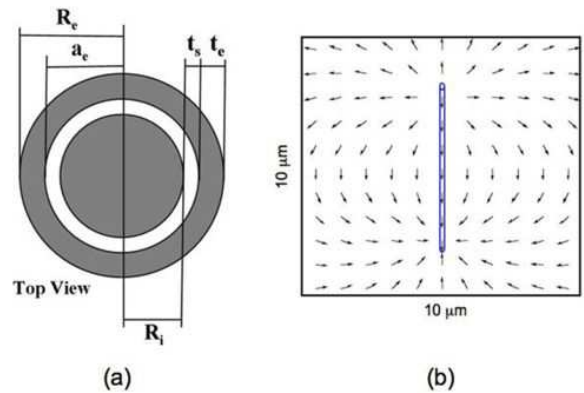}
\end{center}
\caption{(a) Geometrical parameters of the cross section of the multilayer
microwire. (b) Magnetostatic field profile for a tube uniformly magnetized
with $L=6$ mm, $R_{e}=25$ $\protect\mu$m and $a_{e}=20.8$ $\protect\mu$m.}
\end{figure}

Due to the geometry, the magnetization of the complete microwire should be
strongly influenced by the magnetostatic interaction between the nucleus and
the external microtube. As a consequence, a shift of the low-field
hysteresis loop is expected to appear, which is not related to any kind of
exchange \cite{NS99, STZ+03} but rather to magnetostatic biasing effect. The
possibility to achieve such spin-valve-like hysteresis loop is very
attractive for its applications to sense magnetic fields in magnetic
recording systems.

Although the magnetic behavior of multilayer microwires has been intensely
investigated experimentally, \cite{VBK+07, TKP+07, TBP+07} analytical
calculations of the magnetostatic interaction between the magnetic layers
have not yet been performed. Our intention is to fill this gap by proposing
a simple model for the magnetostatic interaction in a multilayer microwire.
From the derived interaction energy we obtain a magnetostatic bias field
that can be directly compared to experimental results. Furthermore the
analytical result is generalized for a broad range of geometrical parameters.

\section{Experimental Details}

Multilayer magnetic microwires have been prepared with suitable composition
and characteristics to obtain optimized soft/hard multilayer system so that
inter-layer magnetostatic interaction can be properly detected. A detailed
description of the fabrication technique can be found elsewhere. \cite%
{ZGV+03, PHN+04} The precursor microwire investigated consists of an
amorphous nucleus ($R_{i}=8.7$ $\mu $m) with nominal composition Co$_{67.06} 
$Fe$_{3.84}$Ni$_{1.44}$B$_{11.53}$Si$_{14.47}$Mo$_{1.66}$. For this alloy
composition a balanced magnetostriction of the order of $10^{-7}$ is
reached, resulting in an ultrasoft magnetic behavior. This soft nucleus is
coated by an intermediate non-magnetic Pyrex layer ($t_{s}=12.1$ $\mu $m
thick) produced by the quenching and drawing method. \cite{ZGV+03} Then, a
30 nm thick Au nanolayer is sputtered onto the Pyrex coating, which plays
the role of electrode/substrate for the final electroplating of the
magnetically harder CoNi outer microlayer. The thickness of the CoNi layer
depends on the current density and the electroplating time, while the
relative amount of Co and Ni slightly varies with the current density. \cite%
{PPG+05} In the present study, the electroplating was performed at 40 $%
^{\circ}$C at fixed current density of 6 mA/cm$^{2}$ during a time that was
varied up to 60 min. A thickness growth ratio of 0.28 $\mu $m/min for the
electroplated shell has been determined by Scanning Electron Microscopy. The
composition of the magnetic external shell was Co$_{90}$Ni$_{10}$, with a
saturation magnetization $M_{e}^{0}=1.2\times 10^{6}$ A/m as determined by
XR fluorescence. A relatively hard magnetic behavior is found for this
external shell.

The magnetic properties have been determined in a VSM and inductive
magnetometer at the temperatures 5 and 300 K with the applied magnetic field
parallel to the microwire axis. Two series of hysteresis loops have been
measured: (\textit{i}) high-field loops ($\pm 40$ kA/m maximum applied
field) to analyze the magnetic behavior of the multilayer microwire, and (%
\textit{ii}) low-field loops ($\pm 1$ kA/m maximum field) measured after
premagnetazing under a nearly saturating field of $\pm 80$ kA/m to obtain
information about the magnetostatic interaction between different layers.
Such magnetostatic coupling has been analyzed as a function of the
dimensions of the hard layer: thickness and length, respectively. All
samples were carefully measured in the same position and orientated with the
North magnetic pole, where the terrestrial field is well known and it is
removed after the measurements.

\section{Theoretical calculations}

Our model system consists of a microtube made of a hard magnetic material
with magnetization saturation $M_{e}^{0}$, containing in its interior a soft
magnetic cylindrical nucleus with saturation magnetization $M_{i}^{0}$. We
adopt a simplified approach in which the discrete distribution of magnetic
moments of both nucleus and tube is replaced by a continuous one, defined by
a function $\mathbf{M}(\mathbf{r})$ such that $\mathbf{M}(\mathbf{r})\delta V
$ gives the total magnetic moment within the element of volume $\delta V$
centered at $\mathbf{r}$. We consider wires for which $L\gg R_{e}$ so it is
reasonable to assume an axial magnetization due to shape anisotropy, defined
by $\mathbf{M}(\mathbf{r})=M_{0}\mathbf{\hat{z}}$, where $\mathbf{\hat{z}}$
is the unit vector parallel to the wire axis. The magnetostatic interaction
between the nucleus and the microtube can be calculated from \cite{Aharoni}%
\begin{equation*}
E_{int}=\mu _{0}\int \mathbf{M}_{i}(\mathbf{r})\nabla U_{e}(\mathbf{r})\,dV\
,
\end{equation*}%
where $\mathbf{M}_{i}(\mathbf{r})$ is the magnetization of the nucleus and $%
U_{e}(\mathbf{r})$ is the magnetostatic potential of the external microtube.
The expression for this potential has been previously reported \cite{EAA+08}
and is given by 
\begin{multline}
U\left( r,z\right) =\frac{M_{e}^{0}}{2}\int_{0}^{\infty }\frac{dk}{k}%
J_{0}\left( kr\right)   \label{potential} \\
\left[ R_{e}J_{1}\left( kR_{e}\right) -a_{e}J_{1}\left( ka_{e}\right) \right]
\\
\left( e^{-k\left\vert \frac{L}{2}-z\right\vert }-e^{-k\left\vert -\frac{L}{2%
}-z\right\vert }\right) 
\end{multline}%
From (\ref{potential}) it is possible to obtain the expression for the
magnetostatic field. Thus we write, $\mathbf{H_{e}}\left( r,z\right)
=-\nabla U_{e}=H_{r}\left( r,z\right) \mathbf{\hat{r}}+H_{z}\left(
r,z\right) \mathbf{\hat{z}}$ with 
\begin{multline*}
H_{r}\left( r,z\right) =\frac{M_{e}^{0}}{2}\int_{0}^{\infty }dk\left[
R_{e}J_{1}\left( kR_{e}\right) -a_{e}J_{1}\left( ka_{e}\right) \right]  \\
J_{1}\left( kr\right) \left( -e^{-k\left\vert \frac{L}{2}-z\right\vert
}+e^{-k\left\vert -\frac{L}{2}-z\right\vert }\right) 
\end{multline*}%
and 
\begin{multline*}
H_{z}\left( r,z\right) =\frac{M_{e}^{0}}{2}\int_{0}^{\infty }dk\left[
R_{e}J_{1}\left( kR_{e}\right) -a_{e}J_{1}\left( ka_{e}\right) \right]  \\
J_{0}\left( kr\right) {\LARGE (}\mbox{sign}\left( \frac{L}{2}-z\right)
e^{-k\left\vert \frac{L}{2}-z\right\vert } \\
-\mbox{sign}\left( -\frac{L}{2}-z\right) e^{-k\left\vert -\frac{L}{2}%
-z\right\vert }{\LARGE )}
\end{multline*}%
Figure 1(b) illustrates the magnetostatic field profile for a tube with $L=6$
mm, $a_{e}=20.8$ $\mu $m and $R_{e}=25$ $\mu $m.

Finally, the magnetostatic interaction between the two magnetic layers in
the microwire is given by%
\begin{multline*}
E_{int}=\pi \mu _{0}M_{i}^{0}M_{e}^{0}R_{i}R_{e}\int\limits_{0}^{\infty }%
\frac{dk}{k^{2}}J_{1}(kR_{i}) \\
\left( J_{1}(kR_{e})-\frac{a_{e}}{R_{e}}J_{1}(ka_{e})\right) \left(
1-e^{-kL}\right) \ ,
\end{multline*}%
where $J_{1}$ is a Bessel function of first kind and first order.

The coercivity, $H_{c}$, of a soft magnetic nucleus in the multilayer
microwire can be calculated as 
\begin{equation*}
H_{c}=H_{c}^{i}-H_{b}\ ,
\end{equation*}%
where $H_{c}^{i}$ is the coercivity of the soft magnetic nucleus if isolated
and $H_{b}$ is the magnetostatic bias field due to the magnetostatic
interaction between the external hard magnetic shell on the soft magnetic
nucleus. This magnetostatic bias field can be written as a function of the
magnetostatic interaction between the magnetic layers%
\begin{equation}
H_{b}=\frac{E_{int}}{\mu _{0}M_{i}^{0}V}\ .  \label{1}
\end{equation}

\section{Results}

Figure 2 shows the high-field hysteresis loop for a multilayer microwire
with $t_{e}=4$ $\mu $m at room temperature. As observed, the magnetization
process takes place in two steps: a giant Barkhausen jump in very low field,
of a few A/m, that must be ascribed to the magnetic amorphous nucleus; and
an additional spread jump (in the range between $10$ and $20$ kA/m),
corresponding to the harder CoNi outer shell. The flat region before the
second step is reversible until a field of $3$ kA/m. At higher fields an
irreversible contribution appears. The fractional magnetization jump and the
effective coercivity of the loop depend on the thickness and composition of
the hard shell. \cite{PPG+05, TBP+07}

\begin{figure}[h]
\begin{center}
\includegraphics[width=8cm]{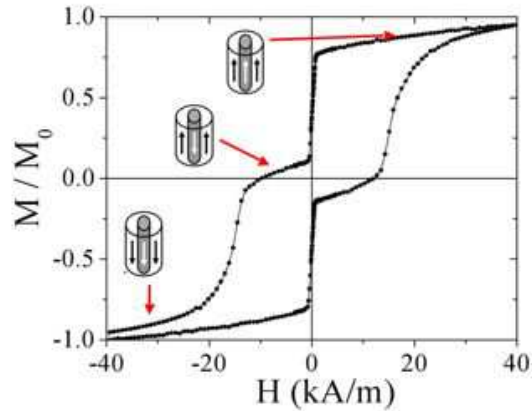}
\end{center}
\caption{High-field hysteresis loop for a soft/hard multilayer microwire
with a thickness of the outer hard magnetic layer $t_{e}=4$ $\protect\mu$m.}
\end{figure}

As observed in Fig. 2, the loop is symmetrical, indicating that no relevant
interaction between magnetic layers is present. To investigate the
magnetostatic biasing effect in the multilayer microwires, the low-field
hysteresis loops of the composite microwire have been analyzed. Previous to
these low-field measurements, the microwires were premagnetized under a
nearly saturating dc magnetic field. The study is then performed under
maximum applied fields smaller than the one required to reverse the
magnetization of the hard outer shell, in order that only the internal
nucleus contributes to the hysteresis loops.

\subsection{Dependence of the thickness of the CoNi microtube on the dipolar
bias}

Figure 3 shows the low-field loop for wires with $L=6$ mm, $R_{i}=8.7$ $\mu$%
m and $t_{s}=12.1$ $\mu$m, considering different values of thickness, $t_{e}$%
, of the outer microtube, measured after premagnetization at $\pm 80$ kA/m.

\begin{figure}[h]
\begin{center}
\includegraphics[width=8cm]{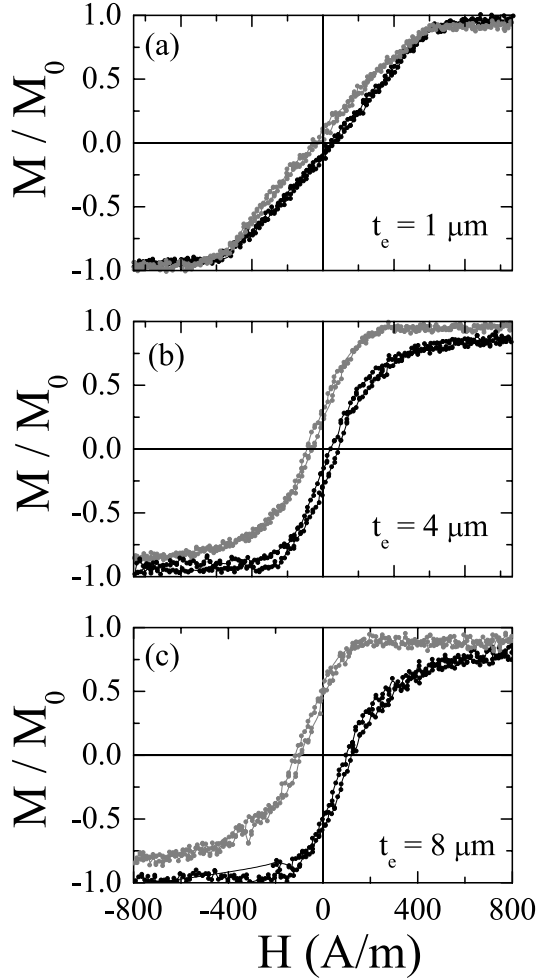}
\end{center}
\caption{Low-field hysteresis loops for composite microwires with $L=6$ mm, $%
R_{i}=8.7$ $\protect\mu $m and $t_{s}=12.1$ $\protect\mu $m, after
premagnetizing at $+80$ kA/m (black dots illustrate both branches of the
cycle) and $-80$ kA/m (gray dots illustrate both branches of the cycle) for
a CoNi thickness of (a) $t_{e}=1$ $\protect\mu $m, (b) $t_{e}=4$ $\protect%
\mu $m and (c) $t_{e}=8$ $\protect\mu $m.}
\end{figure}

It should be emphasized that the premagnetazing process is needed in order
to observe the magnetostatic biasing effect, that is, the shift of the loops
towards the orientation of the premagnetizing field. With this process the
hard layer remains practically at its remanence state, creating a fixed
non-homogeneous magnetic field at the soft nucleus of the multilayer
microwire. The sign of this magnetic field depends on the orientation of the
premagnetizing field. When the outer shell is positively magnetized the
field it produces in the nucleus region is basically negative, thus shifting
the hysteresis loop towards positive values as confirmed in Fig. 3. This
shift is consistent with the definition of the magnetostatic bias field, $%
H_{b}$, and is similar to the one due to exchange coupling in FM/AFM
bilayers. Since exchange coupling can be discarded due to the thickness of
the intermediate non-magnetic layers, the coupling must be of magnetostatic
origin. Moreover, the presence of the premagnetized hard outer shell
introduces another asymmetry behavior in the magnetization curve of the soft
layer, that is, a second magnetization region with lower susceptibility
appears at higher positive or negative field according with positive or
negative premagnetization, respectively. Such asymmetric behavior is due to
the inhomogeneity of the magnetostatic bias field. The magnetic volume of
the second magnetization region increases with the CoNi thickness as shown
in figure 3 in Ref. 14. The previously study of magnetization profiles in
trilayer ribbons with the same magnetic configuration (CoNi/CoFeSiB/CoNi),
has proved that the second magnetization region with low susceptibility is
found close to the uncompensated poles of the hard layer, where the
magnetostatic bias field is stronger. \cite{TKB+08}

Theoretical and experimental magnetostatic results are combined in Fig. 4.
Experimental data for the magnetostatic bias field for different values of
the hard layer thickness are depicted by gray dots and the theoretical
prediction is represented by the solid line. It is clear that there is a
strong dependence of the bias field on $t_{e}$. Note the good agreement
between experimental datapoints and analytical results for multilayer
microwires for $t_{e}<10$ $\mu $m.

\begin{figure}[h]
\begin{center}
\includegraphics[width=8cm]{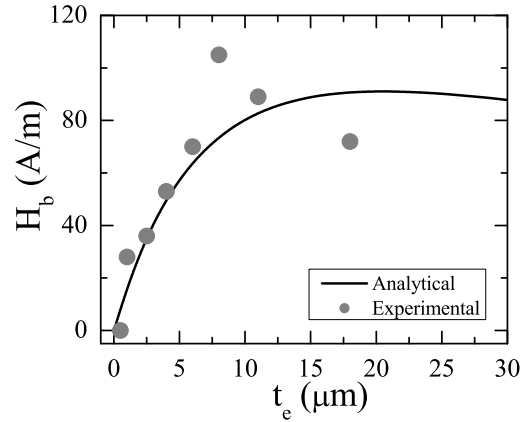}
\end{center}
\caption{Magnetostatic bias field as a function of the thickness of the CoNi
layer. The gray dots correspond to experimental data and the solid line
represents the values calculated analytically. Parameters: $L=6$ mm, $%
R_{i}=8.7$ $\protect\mu$m and $t_{s}=12.1$ $\protect\mu$m. We have used $%
M_{e}^{0}=1.2\times10^{6}$ A/m.}
\end{figure}

\subsection{Dependence of the bias field on the microwire length}

Figure 5 shows the low-field loop for wires of $t_{e}=11$ $\mu $m, $%
R_{i}=8.7 $ $\mu $m and $t_{s}=12.1$ $\mu $m, and different lengths, $L$,
measured after premagnetizing at $\pm 80$ kA/m. The shape of the shifted
loops notably depends on the length of the samples. With increasing length,
the loops become asymmetric and the magnetostatic bias field is reduced. For
longer length, the loops become again symmetric. Further analysis of this
evolution with length is performed elsewhere. \cite{TKP+07, TBP+07}

\begin{figure}[h]
\begin{center}
\includegraphics[width=8cm]{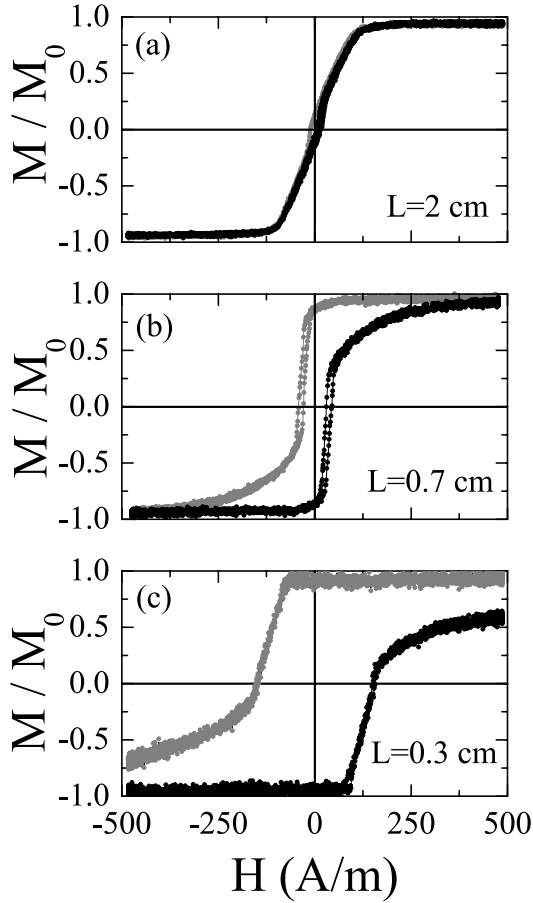}
\end{center}
\caption{Low-field hysteresis loops for composite microwires of $t_{e}=11$ $%
\protect\mu $m, $R_{i}=8.7$ $\protect\mu $m and $t_{s}=12.1$ $\protect\mu $%
m, after premagnetizing at $+80$ kA/m (black dots illustrate both branches
of the cycle) and $-80$ kA/m (gray dots illustrate both branches of the
cycle) for lengths of (a) $L=2$ cm, (b) $L=0.7$ cm and (c) $L=0.3$ cm.}
\end{figure}

Our results are combined in Fig. 6. Experimental data for the magnetostatic
bias field of the system are depicted by gray dots and the solid line
represents the analytical calculation. We observe a strong decrease of the
magnetostatic bias field as the microwire length is increased. The
analytical calculation overestimates the magnetostatic bias field, although
it agrees very reasonably with the experimental results.

\begin{figure}[h]
\begin{center}
\includegraphics[width=8cm]{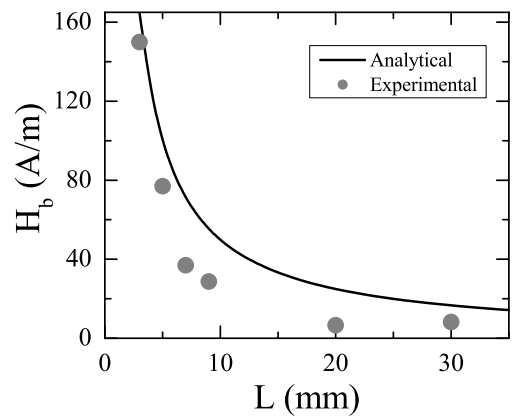}
\end{center}
\caption{Bias field as a function of the length of the multilayer microwire.
The gray dots correspond to the data measured experimentally and the solid
line, the values calculated analytically. Parameters: $t_{e}=11$ $\protect%
\mu $m, $R_{i}=8.7$ $\protect\mu$m and $t_{s}=12.1$ $\protect\mu$m. We have
used $M_{e}^{0}=1.2\times10^{6}$ A/m.}
\end{figure}

\section{Discussion and Conclusion}

The analytical results presented above may be extended to include other
parameter variations. Figure 7 illustrates the magnetostatic bias field as a
function of $t_{e}$ for different values of (a) $R_{i}$, (b) $L$ and (c) the
width of the spacer, $t_{s}$. In the range of parameters considered, we
observe that an increase in $R_{i}$ results in an increase of the
magnetostatic bias field. Furthermore, an increase of the length, $L$,
produces a decrease of the magnetostatic bias field. Finally, increasing the
spacer between the two magnetic layers produces a decrease of the
magnetostatic bias field felt by the inner microwire.

\begin{figure}[h]
\begin{center}
\includegraphics[width=8cm]{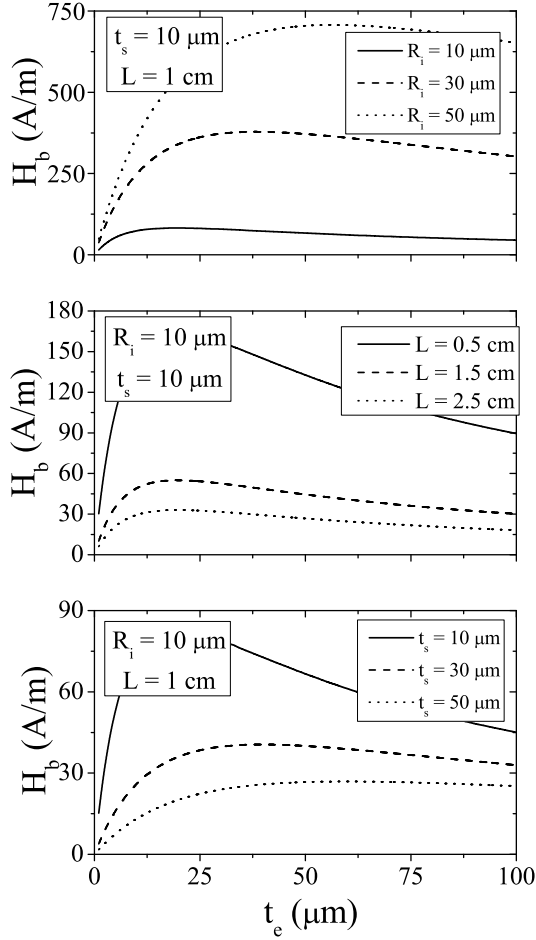}
\end{center}
\caption{Bias field as a function of $t_{e}$ for different values of (a) $%
R_{i}$, (b) $L$ and (c) $t_{s}$. We have used $M_{e}^{0}=1.2\times10^{6}$
A/m.}
\end{figure}

In summary, we have investigated the magnetostatic biasing effect in
multilayer microwires. Using a continuous model we have obtained a simple
expression to model the magnetostatic bias in these particles. From our
calculations and measurements we can conclude that the magnetostatic bias
field strongly depends on the geometry of the system, achieving large
values, comparable to the coercivity in some cases. Our results provide
guidelines for the production of microstructures with tailored magnetic
properties.

\section{Acknowledgments}

This work was supported by Fondecyt (No. 11070010 and 1080300), Millennium
Science Nucleus \textit{Basic and Applied Magnetism} (project P06-022F), and
Spanish MICINN under research project MAT2007-65420-C02-01. M. Bahiana
acknowledges support from Instituto do Milenio de Nanotechnologia, MCT/CNPq,
FAPERJ, and PROSUL/CNPq.

\end{document}